\documentclass[12pt,a4paper,english,superscriptaddress,aps,preprint]{revtex4}
\usepackage{amsmath}
\usepackage{amssymb}
\usepackage{graphicx}
\usepackage{hhline}
\usepackage{mwe}
\usepackage{textcomp}
\makeatletter
\usepackage{babel}
\newcommand{\bea}{\begin{eqnarray}}
\newcommand{\eea}{\end{eqnarray}}

\newcommand{\pa}{\partial}

\newcommand{\be}{\begin{equation}}
\newcommand{\ee}{\end{equation}}
\numberwithin{equation}{section}

\begin{document}
\immediate\write16{<<WARNING: LINEDRAW macros work with emTeX-dvivers
                    and other drivers supporting emTeX \special's
                    (dviscr, dvihplj, dvidot, dvips, dviwin, etc.) >>}

\title{\boldmath Particle propagation on spacetime manifolds with static defects}
\preprint{KA-TP-04-2017}

\author{Jose M. Queiruga}
\affiliation{Institute for Theoretical Physics, Karlsruhe Institute
of Technology (KIT), 76131 Karlsruhe, Germany}
\affiliation{Institute for Nuclear Physics, Karlsruhe Institute of Technology, Hermann-von-Helmholtz-Platz 1,
D-76344 Eggenstein-Leopoldshafen, Germany}

\email{jose.queiruga@kit.edu}

\begin{abstract}
We investigate the effects of small static defects in the spacetime manifold. The presence of the defects leads to a modification of the scalar field two-point function in Klein-Gordon theory. We calculate the energy-momentum tensor and discuss the possible mass generation for the scalar field in single and multiple defect spacetimes. We also extend these results to the photon field and show that, as a result of the interaction with the defects, the photon dispersion relations are modified.

\end{abstract}

\maketitle


\section{Introduction}
\label{intro}

Half a century ago, it was pointed out by Wheeler \cite{Wheeler1,Wheeler2}, that at lengths comparable with the Planck length scale, one should expect large fluctuations of the metric and topology. These fluctuations modify the short-distance structure of spacetime, which can be thought of as a sort of foam of spacetime. A few years later, Hawking \cite{Hawking1}  introduced a mathematical framework to describe  this spacetime foam in terms of a path integral approach to favor the treatment of nontrivial topologies. Within this framework it is even possible to perform the path integration over a certain class of metrics and topologies.

If the spacetime has this foam-like structure at the Planck scale, one should expect the propagation of particles to be affected, even for flat metrics and nontrivial topological structure. If we think of the spacetime foam as a gas of defects one may expect of course, that it must exist throughout spacetime, but for purposes of interpretation, it is useful to assume that it is concentrated over a compact region, such that spacetime is asymptotically flat and topologically trivial very far from it. In  \cite{Hawking2,Hawking3} the authors consider three different defect topologies, namely $\mathbb{C}P^2$,  $\mathbb{S}^2\times\mathbb{S}^2$ and $K3$. These four-dimensional defects allow for the treatment of single-defect manifolds. However, results regarding spacetimes made of an infinite number of such defects seem difficult to obtain. Other approaches to the spacetime foam are based on stringy analogues. In  \cite{Mav1} the photon is modeled as an open string  and the foam as a gas of D-particles. The interaction of the photons with the D-particles leads to a nontrivial vacuum refractive index and therefore, to a modification of the dispersion relations. A different approach was presented in \cite{Klink1}. In this paper a classical treatment of three types of static defect-manifolds was performed (the defects are generated after removing a three-sphere from the spatial section of the spacetime and imposing different identification conditions to the points on the surface). The classical solutions of the Maxwell equations, modified properly to accommodate the topology of the different defect manifolds, allow nonstandard modified dispersion relations for the plane-wave solutions to be obtained. Similar results were obtained in \cite{Klink2, Klink22} for line defects. This is the approach we will take here.

This point of view, defining the spacetime foam as a set of static defects in a spatial slice of the usual Minkowski spacetime, bears some similarities with the Casimir effect set-up \cite{Casimir}. In the latter, space is modified by including regions where the fields must obey certain boundary conditions, breaking for example, the translational invariance. As a consequence of this modification the vacuum energy is shifted and the propagation of particles affected. In the case of two infinite parallel perfectly conducting plates at finite distance, the photon dispersion relations are modified (this behavior appears at two-loop order) in the direction perpendicular to the plates \cite{Schan}. Similar results were obtained when the spacetime possesses nontrivial topological structure \cite{Isham,Unwin,Dowker}. In these examples, the modifications in the spacetime are ``global" in the sense that, there is no asymptotic Minkowski region. In our defect manifolds, the defects are ``localized" in the sense that it is possible to define a distance to the defect and physical quantities such as vacuum energy or generated mass will recover their usual (Minkowski) values sufficiently far from the defect. Of course, when the defect size tends to zero, our defect manifolds collapse to the Minkowski spacetime, and therefore all physical quantities will approach their usual values again. We will also consider spacetimes with an infinite number of defects. In this case, the physical quantities do not depend on the spacetime coordinates but on the parameters characterizing the distribution: the size of the defects and the density of the gas. Since there is no asymptotical Minkowski region the usual Lorentz-invariant results are recovered only in the limit of vanishing gas density and/or defect size. 

This paper is organized as follows. In Sec. \ref{def}, we define the defect manifolds and discuss some features of the solutions of the massless Klein-Gordon equation for different spacetime topologies. In Sec. \ref{greense}, we present the general structure of the Green functions due to the presence of the defects. In Secs. \ref{tauzero} and \ref{tauone}, the Green functions in $\mathcal{M}^{\tau=0}_4$ and $\mathcal{M}^{\tau=1}_4$  are calculated. With this Green function we calculate the vacuum stress-energy tensor and the generated mass in $\lambda\phi^4$ theory. In Sec. \ref{pois} we discuss the relation between Poisson processes, Lorentz symmetry and the gas of defects for scalar particles. In Sec. 7 we extend these results to the photon field. Finally, in Sec. \ref{sum}, we summarize and discuss our results. We also add an appendix with useful integrals.


\section{Defect manifolds}
\label{def}

Throughout this paper we will consider a variety of spacetimes of the form $\mathcal{M}_4^{(\tau)}=\mathbb{R}\times \mathcal{M}^{(\tau)}$ \cite{Klink1,Klink3}, where the spatial sections $ \mathcal{M}^{(\tau)}$ are three-dimensional manifolds constructed in one of the following two ways
\bea
\mathcal{M}^{(\tau=0)}&=&\lbrace \vec{x}\in\mathbb{R}^3: \left(\vert \vec x \vert \geq b>0 \right) \rbrace,\label{M1}\\
\mathcal{M}^{(\tau=1)}&=&\lbrace \vec{x}\in\mathbb{R}^3: \left(\vert \vec x \vert \geq b>0 \right)\wedge\left(P (\vec{x})\cong\vec{x}\,\, \text{for}\,\, \vert \vec{x}\vert=b \right)  \rbrace\label{M2},
\eea
where
\bea
P((x,y,z))&=&-(x,y,z)\label{ide1}.
\eea

The topological structure of these manifolds can be seen by means of the inversion diffeomorphism
\be
r\rightarrow\rho=\frac{b}{r},
\ee
where $r=\vert\vec{x}\vert$. We get 
\be
\mathbb{R}\times\mathcal{M}^{(\tau=0)}\simeq \mathbb{R}\times\left(\mathbb{D}^3\setminus \{p\}\right)\quad \text{and}\quad \mathbb{R}\times\mathcal{M}^{(\tau=1)}\simeq \mathbb{R}\times\left(\mathbb{R}P^3\setminus \{p\}\right),
\ee
(the point $p$ corresponds to the spatial infinity in the inversion).

\begin{figure}[h]       
    \fbox{\includegraphics[width=0.3\textwidth]{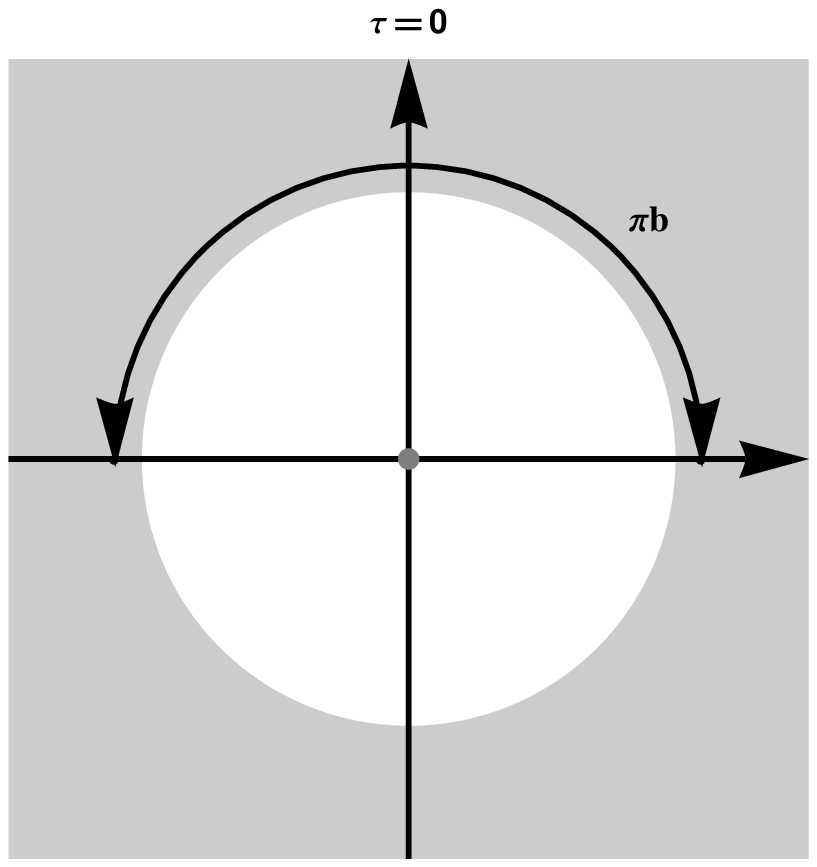}}   
    \hspace{30px}
    \fbox{\includegraphics[width=0.3\textwidth]{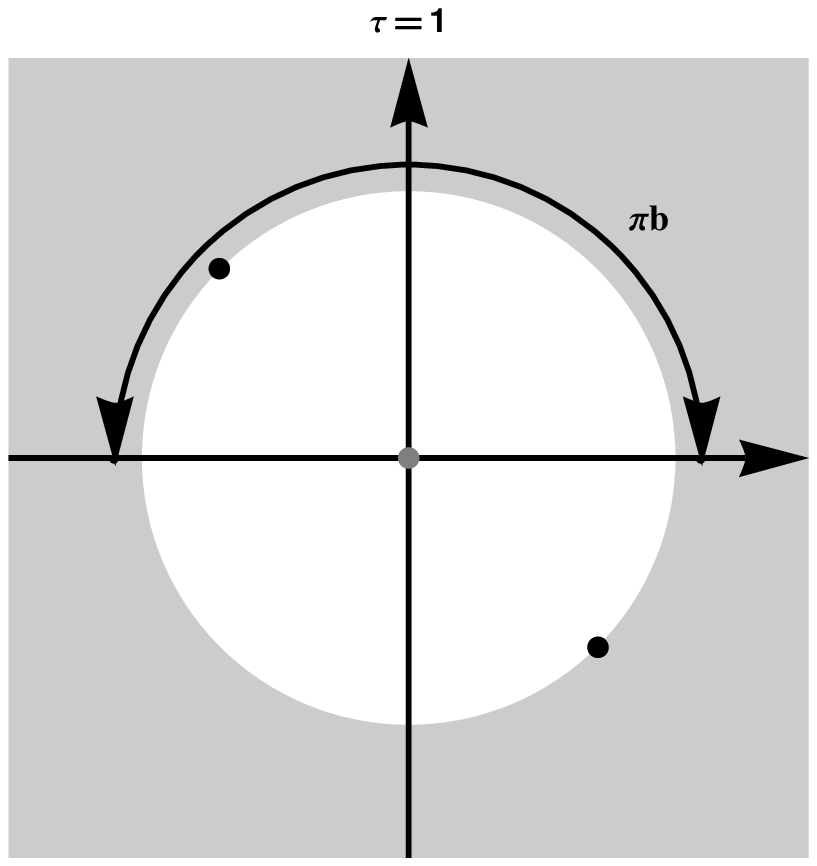}}
    \hspace{30px}
    \caption{Two dimensional representation of the manifolds $\mathcal{M}^{(\tau)}$ after the diffeomorphism inversion. The black points represent the identification.}
    \label{materialflowChart}
\end{figure}

The first type ($\tau=0$) was constructed simply by removing a three-disk ($\mathbb{D}^3$) of radius $b$ from the usual Minkowski spacetime while for $\tau=1$ in addition we identify the points at the boundary of the three-disk as indicated in (\ref{ide1}). It is important to note that, in the $\tau=0$ case the manifold has a boundary corresponding to the defect surface. In the usual definition of spacetime defects the boundaries are removed after certain identification of the points in the surface (e.g. the $\tau=1$ case). However, for computational purposes we will start here with the $\tau=0$ case as a preliminary for the $\tau=1$. Since the defects are spherical it is natural to work in spherical coordinates. In these coordinates the identification condition can be rewritten as follows
\be
P((r,\theta,\varphi))=(r,\pi-\theta,\pi+\varphi)\label{id1s}.
\ee

Of course $\tau=0,1$ manifolds are topologically different. For example, regarding the homotopical structure they differ in the first homotopy group. For $\tau=0$ it is trivial, while for $\mathbb{R}P^3$ it is $\mathbb{Z}_2$ (the higher homotopy groups coincide). With these building blocks we can construct more complicated structures modeling the spacetime foam by intersecting the simple structures (\ref{M1}) and (\ref{M2}), 
\be
\mathcal{M}^{(\tau)}_{\{\vec{x}_1,...,\vec{x}_N\}}=\bigcap_{i=1}^N \mathcal{M}_{\vec{x}_i}^{(\tau)}\label{mani11}
\ee
where the points $\vec{x}_i$ denote the center of the removed spheres and we can take eventually $N\rightarrow\infty$. An adequate distribution of the points $\vec{x}_i$ (we will explain this later) allows us to define a model for the spacetime foam based on (\ref{mani11}). As an illustrative example let us study the classical solutions in massless Klein-Gordon theory. They are modified in order to satisfy the different boundary conditions. For $\mathcal{M}^{\tau=0}$ we get (with $\omega^2=k^2$)
\be
\phi^{(\tau=0)}(t,\vec{x})=\sum_{l,m}\int d k A^{(0)}_{k,l,m}\left(j_l(kr)-\frac{j_l(kb)}{y_l(kb)}y_l(kr)\right)Y_l^m e^{-i\omega t}\label{t0},
\ee
where $j_l(kr)$, $y_l(kr)$ and $Y_l^m$  are the spherical Bessel functions of first and second kind and the spherical harmonics respectively and we have imposed $\phi^{(\tau=0)}(t,\vec{x})\vert_{\pa B_b}=0$ ($\pa B_b$ is the defect surface). For the $\tau=1$ defect we obtain
\bea
\phi^{(\tau=1)}(t,\vec{x})&=&\sum_{l\, odd,m}\int d k A^{(1)}_{k,l,m}\left(j_l(kr)-\frac{j_l(kb)}{y_l(kb)}y_l(kr)\right)Y_l^m e^{-i\omega t}+\nonumber\\
&&\sum_{l\, even,m}\int d k \left(A^{(1)}_{k,l,m}j_l(kr)-B^{(1)}_{k,l,m}y_l(kr)\right)Y_l^m e^{-i\omega t}\label{t1},
\eea
where we have imposed $\phi^{(\tau=1)}(t,\vec{x})\vert_{\pa B_b}=\phi^{(\tau=1)}(t,-\vec{x})\vert_{\pa B_b}$ (note that $Y_l^m(\pi-\theta,\pi+\varphi)=(-1)^l Y_l^m(\theta,\varphi)$).

 It is interesting to note that, unlike the $\tau=0$ case, where the set of solutions is completely determined up to an overall constant for each mode $A^{(0)}_{k,l,m}$, in the case $\tau=1$ we have three sets of constants: $A^{(1)}_{k,l,m}$ for odd $l$ and $A^{(1)}_{k,l,m},B^{(1)}_{k,l,m}$ for even $l$. These constants can be fixed, for example, by continuity of the first derivative on the identified points and the Sommerfeld radiation condition \cite{Atkinson} (we require that the solution must behave as $e^{ikr}/r$ at $r=\infty$). We will analyze the behavior of the Green functions of the D'Alembertian operator through the following sections.


\section{Two-point Green function in nontrivial spacetime}
\label{greense}
Let us start with a massless scalar field in four dimensions obeying the massless Klein-Gordon equation
\be
\square \phi^{(\tau)}=0,
\ee
where the superscript $\tau$ stands for the two different defects. The two-point Green function can be defined as follows
\be
G(x,x')=-i\langle 0\vert T \phi^{(\tau)}(x)\phi^{(\tau)}(x')\vert 0\rangle,
\ee
where $x=(t,\vec{x})$. We can obtain the Green function from the expansion in modes of the solution of the Klein-Gordon equation with the appropriate boundary conditions (\ref{t0})-(\ref{t1})  or equivalently by solving the inhomogeneous equation
\be
\square G(x,x')=-\delta^{(3)} (\vec{x}-\vec{x}')\delta(t-t').\label{green0}
\ee

Since we are interested in static defects it is natural to take the Fourier transform on the time coordinate
\be
G_\omega(\vec{x},\vec{x}')=\int dt e^{-i\omega (t-t')}G(x,x').
\ee

This new Green function satisfies the following equation
\be
\left( \omega^2 +\nabla^2  \right)G_\omega(\vec{x},\vec{x}')=\delta^{(3)}(\vec{x}-\vec{x}') \label{eqgreen}.
\ee

Before imposing boundary conditions on the surface defect (we impose the Sommerfeld radiation condition at $r=\infty$) it can be shown \cite{Bender} that the general solution of (\ref{eqgreen}) has the form
\be
G_\omega(r,r',\gamma)=\sum_{n=0}^\infty a_n \frac{1}{\sqrt{r}}H^{(1)}_{n+1/2}\left(\vert \omega\vert r\right)P_n(\cos\gamma), \quad r>r'\label{g1}
\ee
and
\be
G_\omega(r,r',\gamma)=\sum_{n=0}^\infty  \frac{1}{\sqrt{r}}\left(b_n H^{(1)}_{n+1/2}\left(\vert \omega\vert r\right)+c_n H^{(2)}_{n+1/2}\left(\vert \omega\vert r\right)\right)P_n(\cos\gamma), \quad r'>r\label{g2}
\ee
where
\be
\cos\gamma=\sin \theta  \sin \theta ' \cos \left(\phi -\phi
   '\right)+\cos \theta  \cos \theta '
\ee
and $H_{n+1/2}^{(i)},\, i=1,2$ are the Hankel functions of the first and second kind. The coefficients $a_n,b_n$ and $c_n$ are determined by three conditions, namely:\\

1. The boundary conditions over the defect surface
\be
B^{(\tau)}\left[G_\omega(r,r',\gamma)\right]\vert_{r=b}=0\label{b1}.
\ee

2. The continuity at $r=r'$,
\be
b_n H^{(1)}_{n+1/2}\left(\vert \omega\vert r'\right)+c_n H^{(2)}_{n+1/2}\left(\vert \omega\vert r'\right)=a_n H^{(1)}_{n+1/2}\left(\vert \omega\vert r'\right)\label{b2}.
\ee

3. The jump condition of the first derivative at $r=r'$ (see appendix B),
\be
 b_n\pa_r H^{(1)}_{n+1/2}\left(\vert \omega\vert r\right)\vert_{r=r'}+c_n\pa_r H^{(2)}_{n+1/2}\left(\vert \omega\vert r'\right)\vert_{r=r'}-a_n \pa_rH^{(1)}_{n+1/2}\left(\vert \omega\vert r'\right)\vert_{r=r'}=\frac{2n+1}{4\pi r'^{3/2}}\label{b3}.
\ee

The second and third conditions do not depend on the particular topology of the defect, i.e. on the identification of the points at the boundary. The topological information is completely encoded in the first condition. As we will see below, $B^{(\tau)}$ depends on the type of defects and it is defined explicitly in Eqs. (\ref{bb0}) and (\ref{bb1}) . We will determine the different Green functions for each manifold in the following sections.


\section{Green function: $\mathcal{M}_4^{(\tau=0)}$ manifold}
\label{tauzero}

Let us start with the defect manifold $\mathcal{M}_4^{(\tau=0)}$. In this case, we simply remove a ball of radius $b$ centered at the origin. The boundary condition (\ref{b1}) will be
\be
B^{(\tau=0)}\left[G_\omega(r,r',\gamma)\right]\vert_{r=b}\equiv G_\omega(b,r',\gamma)=0.\label{bb0}
\ee

After imposing (\ref{b2}), (\ref{b3}) and (\ref{bb0}) the coefficients are completely determined:
\bea
c_n&=&\frac{i(2n+1)}{16 \sqrt{r'}}H^{(1)}_{n+1/2}(r'\vert \omega\vert)\label{coe1},\\
b_n&=& -\frac{H^{(2)}_{n+1/2}(b\vert \omega\vert)}{H^{(1)}_{n+1/2}(b\vert \omega\vert)}c_n\label{coe2},\\
a_n&=&b_n+\frac{H^{(2)}_{n+1/2}(r'\vert \omega\vert)}{H^{(1)}_{n+1/2}(r'\vert \omega\vert)}c_n\label{coe3}.
\eea

An observation is in order. Since $\lim_{b\rightarrow 0}\mathcal{M}_4^{(\tau=0)} =\mathcal{M}_4$, i.e. when the size of the defect tends to zero, the defect manifold tends to Minkowski spacetime,  we must have
\be
\lim_{b\rightarrow 0} G_\omega (r,r',\gamma)=G_{\omega,\text{free}} (r,r',\gamma),
\ee
where $G_{\omega,free} (r,r',\gamma)$ is the usual Green function in Minkowski spacetime. In order to show this, we have from (\ref{g1}), (\ref{g2}) and (\ref{coe1})-(\ref{coe3}) the following
\be
\lim_{b\rightarrow0}G_\omega (r,r',\gamma)=\sum_{n=0}^\infty i\frac{2n+1}{8\sqrt{r r'}}H^{(1)}_{n+1/2}(r'\vert \omega \vert)J_{n+1/2}(r\vert \omega \vert)P_n(\cos\gamma).\label{lim1}
\ee

Now we use the identity
\be
\frac{H_\nu^{(1)} (x)}{x^\nu}=2^\nu \Gamma(\nu)\sum_{n=0}^\infty (\nu+n)\frac{H_{\nu+n}^{(1)}(u)}{u^\nu}\frac{J_{\nu+n}(v)}{v^\nu}C_k^{(\nu)}(\cos\gamma)\label{id1},
\ee
where $x=\sqrt{u^2+v^2-2 u v \cos\gamma}$ and $C_k^{(\nu)}$ are the Gegenbauer polynomials (with $C_k^{(1/2)}(z)=P_n(z)$ the Legendre polynomials), see for example \cite{Gradshtein}. From (\ref{lim1}) and (\ref{id1}) we obtain finally
\be
G_{\omega,\text{free}} (\vec{x},\vec{x}')=\lim_{b\rightarrow0}G_\omega (\vec{x},\vec{x}')=\frac{e^{i\vert\omega\vert \vert \vec{x}-\vec{x}'\vert}}{4\pi\vert \vec{x}-\vec{x}'\vert}\label{p1},
\ee
which is our desired result. After taking the inverse Fourier transform in the frequency parameter we can write (\ref{p1}) as follows
\bea
G_\text{free}(x,x')=\frac{i}{4\pi^2\left(x-x'\right)^2}.
\eea

The first obvious observation derived from this fact, is that the Green function will be written as a sum of the free Green function, plus a correction due to the presence of the defect. Let us analyze the Green function (\ref{g2}). Since there is no $b$ dependence in the set of coefficients $c_n$, they only contribute to the free part of the Green function. Therefore, the correction function is completely determined by the expansion of the $b$-dependent part in (\ref{coe2})
\be
-\frac{H^{(2)}_{n+1/2}(b\vert \omega\vert)}{H^{(1)}_{n+1/2}(b\vert \omega\vert)}=-1+\frac{4}{2+\frac{i \pi  4^{-n} (\vert b \omega\vert)^{2 n+1}}{\Gamma
   \left(n+\frac{1}{2}\right) \Gamma \left(n+\frac{3}{2}\right)}}+\mathcal{O}\left((b\vert\omega\vert)^{2n+2}\right).\label{exp1}
\ee

We see from (\ref{exp1}) that the first two contributions in the $b \vert\omega \vert$-expansion are determined by the $b_0$ coefficient, hence we can ignore  the coefficients $b_n, \, n\geq 1$ up to order $\left(\vert\omega\vert b\right)^3$. Finally,  we expand for $b \vert\omega \vert$ small
\be
G_\omega (r,r',\gamma)=G_{\omega,\text{free}} (r,r',\gamma)+\frac{b\vert \omega\vert^2}{4\pi}h_0^{(1)}(r\vert \omega\vert) h_0^{(1)}(r'\vert\omega\vert)+\mathcal{O}\left((\vert \omega\vert b)^2\right)\label{twot0},
\ee
where $h_0^{(1)}$ are the spherical Hankel functions of the first kind. Inverting the Fourier transform
\be
G (x,x')=G_{\text{free}} (x,x')-i\frac{b (r+r')}{4 \pi ^2 r r'
   \left((r+r')^2-(t-t')^2\right)}+....\label{green}
\ee

While the free Green function has a pole in the coincidence limit when $x\rightarrow x'$, the correction terms are regular. This implies that in the calculation of the VEV of the energy-momentum tensor, after the subtraction  of the free part, the result will be finite. This modification of the singular structure of the Green function due to the nontrivial topology gives rise to different effects in the particle propagation \cite{Hawking2,Hawking3}.


\subsection{Vacuum energy-momentum tensor}
\label{4.1}

We start with the following Lagrange density in curved spacetime
\be
\mathcal{L}=\frac{1}{2}\sqrt{-g}\left[ g^{\mu\nu}\pa_\mu\phi\pa_\nu\phi-\xi R \phi^2  \right]\label{conf},
\ee
where $R$ is the Ricci scalar and $\xi$ a numerical parameter. The choice $\xi=1/6$ corresponds to the conformally invariant limit. Now taking the planar limit $g^{\mu\nu}\rightarrow\eta^{\mu\nu}$ (and $R\rightarrow0$)  for this value of $\xi$ we obtain the following stress energy tensor (see for example \cite{Blau})
\be
T_{\mu\nu}=\frac{2}{3}\phi_{,\mu}\phi_{,\nu}-\frac{1}{6}\eta_{\mu\nu}\eta^{\sigma\rho}\phi_{,\sigma}\phi_{,\rho}-\frac{1}{3}\phi\phi_{,\mu\nu}+\frac{1}{3}\eta_{\mu\nu}\phi\square\phi.
\ee

This tensor retains some information of the curvature term in the flat limit and it is automatically traceless on-shell $T^\mu_{\,\,\,\mu}=0$. This is not very surprising since $T_{\mu\nu}$ is obtained from (\ref{conf}) by varying the metric $g_{\mu\nu}$ and at the end of the calculation some information about the last term in (\ref{conf}) remains \cite{Birrel,Chernikov,Callan}. We can determine the VEV of the energy-momentum tensor by subtracting the free contribution
\be
\langle  T_{\mu\nu}\rangle=i\lim_{x'^{\mu}\rightarrow x^\mu} \left(\frac{2}{3}\pa_\mu\pa_{\nu'} -\frac{1}{6}\eta_{\mu\nu}\eta^{\sigma\rho'}\pa_{\sigma}\pa_{\rho'}-\frac{1}{3}\pa_{\mu}\pa_{\nu}+\frac{1}{3}\eta_{\mu\nu}\pa_{\rho'}\pa^{\rho'} \right)\left(G (x,x')-G_{\text{free}} (x,x')\right).
\ee

With this definition the Minkowski vacuum has a vanishing energy, and the nontrivial contribution to $T_{\mu\nu}$ comes from the correction to the Green function. We obtain the following result 
\bea
\langle  T_{\mu\nu}\rangle&=&-\frac{b^2}{32\pi^2r^6}k_{\mu\nu}+\frac{b^2}{32\pi^2 r^8}\omega_{\rho\mu}\omega_{\sigma\nu}x^\rho x^\sigma,\label{tensor0}\\
k_{\mu\nu}&=&\text{diag}\left(1,\frac{2}{3},\frac{2}{3},\frac{2}{3}\right),\\
\omega_{\mu\nu}&=&\text{diag}\left(0,1,1,1\right).
\eea

We see from expression (\ref{tensor0}) that the $\langle T_{\mu\nu}\rangle$ does not receive contributions at linear order in $b$ and $\langle T^\mu_{\,\,\,\mu} \rangle=0$. Very far from the defect the energy-momentum tensor tends to zero. This behavior was expected, since when $r\rightarrow \infty$ the field does not feel the defect, and therefore, everything must approach the Minkowski result.


\subsection{Mass generation}
\label{4.2}

It is well-known that self-interaction terms in non-Minkowskian spacetimes can give nonzero mass to the scalar particles. In  \cite{Ford} it was shown that, in the case of $\lambda \phi^4$ interaction and the multiply-connected spacetime $\mathbb{R}\times \mathbb{R}^2\times \mathbb{S}^1$, the scalar particle acquires a mass of order $\lambda/R^2$, where $R$ the radius of $\mathbb{S}^1$.  We will show in this section that the same effect takes place in the manifold $\mathcal{M}_4^{\tau=0}$. In this case, the mass will be proportional to the size of the defect, but it will also depend on the distance to the defect. We start by considering the following theory
\be
\mathcal{L}=\frac{1}{2}\pa_\mu \phi \pa^\mu\phi-\lambda \phi^4.
\ee

\begin{figure}[h]
    \centering
    \includegraphics[width=0.35\textwidth]{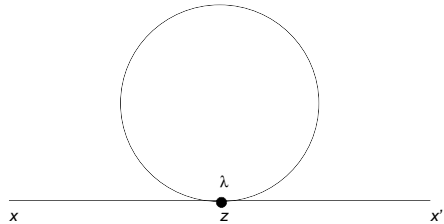}
    \caption{One-loop correction}
    \label{fig1}
\end{figure}

Let us expand the Green function up to first order in the coupling $\lambda$
\be
G(x,x')=G^{(0)}(x,x')-\lambda G^{(1)}(x,x'),
\ee 
where $G^{(0)}(x,x')$ corresponds to expression (\ref{green}) which will be regularized by substracting the free part at the end of the calculation (note that in the theory in Minkowski space the renormalized mass is zero), and $G^{(1)}(x,x')$ is given by the expression (Fig. \ref{fig1})
\be
G^{(1)}(x,x')=-\frac{i}{2}\int d^4z G^{(0)}(x,z)G^{(0)}(z,z)G^{(0)}(z,x').\label{1loop}
\ee

Now if the field $\phi$ acquires mass, the following equation must hold at one-loop order
\be
\left(\square_x+m^2_{\text{eff}}\right)\left(G^{(0)}(x,x')-\lambda G^{(1)}(x,x')\right)=-\delta^{(3)}(\vec{x}-\vec{x}')\delta(t-t'),
\ee 
where $m_\text{eff}^2\propto \lambda$. If we expand and take into account (\ref{green0}) the previous expression can be written as follows
\be
-\delta^{(3)}(\vec{x}-\vec{x}')\delta(t-t')-\lambda\square_x G^{(1)}(x,x')+m^2_{\text{eff}}G^{(0)}(x,x')+\mathcal{O}(\lambda^2)=-\delta^{(3)}(\vec{x}-\vec{x}')\delta(t-t'),
\ee
and finally from (\ref{1loop}) we get the following coordinate dependent mass
\be
m_{\text{eff}}^2=i\frac{\lambda}{2}G^{(0)}(x,x)=\frac{\lambda b}{16\pi^2}\frac{1}{r^3}+\mathcal{O}(b^2).\label{mass1}
\ee


\section{Green function: $\mathcal{M}_4^{(\tau=1)}$ manifold}
\label{tauone}

Our next step is the generalization of the results of the previous sections to the spacetime  $\mathcal{M}_4^{(\tau=1)}$. In order to construct  $\mathcal{M}_4^{(\tau=1)}$ we need to identify the points at the defect surface as indicated in (\ref{id1s}) and therefore, the same identification must be imposed to the Green function. When performing the calculation we need only to take into account the behavior of the Legendre polynomials under parity
\be
P_n(-z)=(-1)^nP_n(z)\label{Leg}
\ee
and $\cos\gamma\rightarrow-\cos\gamma$. As consequence, the boundary conditions (\ref{bb0}) can be written as follows
\be
B^{(\tau=1)}\lbrack G_\omega(r,r',\cos\gamma) \rbrack\vert_{r=b}\equiv G_\omega(b,r',\cos\gamma)-G_\omega(b,r',-\cos\gamma)=0\label{bb1}
\ee
 The Green function splits into two parts. If the index $n$ is odd, we obtain the coefficient of Sec. 4
\bea
c_n&=&\frac{i(2n+1)}{16 \sqrt{r'}}H^{(1)}_{n+1/2}(r'\vert \omega\vert),\\
b_n&=& -\frac{H^{(2)}_{n+1/2}(b\vert \omega\vert)}{H^{(1)}_{n+1/2}(b\vert \omega\vert)}c_n,\\
a_n&=&b_n+\frac{H^{(2)}_{n+1/2}(r'\vert \omega\vert)}{H^{(1)}_{n+1/2}(r'\vert \omega\vert)}c_n.
\eea

  For $n$ even we obtain
 \bea
 c_n&=&\frac{i(2n+1)}{16\sqrt{ r'}}H^{(1)}_{n+1/2}(r'\vert \omega\vert),\\
b_n&=&-\frac{i(2n+1)}{16 \sqrt{r'}}H^{(2)}_{n+1/2}(r'\vert \omega\vert)+a_n.
 \eea

 We still have a free choice for the $a_n$ coefficients. They must be fixed taking into account that in the limit when $b\rightarrow 0$ we want to recover the free Minkowski Green function. But considering (\ref{Leg}), one might expect that the even terms cannot be affected by the boundary conditions. This implies in particular that they cannot depend on the parameter $b$, leading to
 \be
 a_n=c_n\left(1+\frac{H_{n+1/2}^{(2)}(r'\vert \omega\vert)}{H_{n+1/2}^{(1)}(r'\vert \omega\vert)}\right).
 \ee
 
 With this choice, the Green function can be expanded as follows
 \be
G_\omega (r,r',\gamma)=G_{\omega,free} (r,r',\gamma)+\frac{b^3\vert\omega\vert^4}{4\pi}h_1^{(1)}(r\vert\omega\vert)h_1^{(1)}(r'\vert\omega\vert )\cos\gamma  +\mathcal{O}\left((b\vert\omega\vert)^4\right).
\ee

Inverting the Fourier transform produces
\be
G (x,x',\gamma)=G_{free} (x,x',\gamma)-\frac{i b^3 \left(\left(r'+r\right)^3 \left(3 r
   r'+r'^2+r^2\right)-\left(r'^3+r^3\right
   ) \left(t-t'\right)^2\right)}{2 \pi ^2 r^2 r'^2
   \left(\left(r'+r\right)^2-\left(t-t'\right)^2\right)^3}\cos\gamma+...
\ee

Repeating the procedure of Sec. 4  we get
\bea
\langle  T_{\mu\nu}\rangle&=&-\frac{b^3}{12\pi^2r^7}k_{\mu\nu}+\frac{7b^3}{96\pi^2 r^9}\omega_{\rho\mu}\omega_{\sigma\nu}x^\rho x^\sigma,\label{tensor1}\\
k_{\mu\nu}&=&\text{diag}\left(1,\frac{5}{8},\frac{5}{8},\frac{5}{8}\right),\\
\omega_{\mu\nu}&=&\text{diag}\left(0,1,1,1\right).
\eea

 Note that again $\langle T^\mu_{\,\,\,\,\mu}\rangle=0$. For the effective mass due to the $\lambda \phi^4$ interaction we obtain
\be
m_{\text{eff}}^2=\frac{5\lambda b^3}{32 \pi^2}\frac{1}{r^5}+\mathcal{O}(b^4)\label{mass2}
\ee

The Green function in the $\mathcal{M}_4^{(\tau=1)}$ manifold does not have a linear term in $b$. As as consequence the energy-momentum tensor and the mass decay by a $1/r$ and $1/r^2$  factors faster than in the  $\mathcal{M}_4^{(\tau=0)}$ case.


\section{Gas of defects and Lorentz symmetry: Scalar particles}
\label{pois}

In Secs. \ref{tauzero} and \ref{tauone} we have analyzed the particle propagation in single-defect spacetimes. This structure does not provide a good description of the spacetime foam. One might expect instead, that the defects are distributed throughout Minkowski space and not located at a single point. But the explicit calculation of the Green functions in such a general spacetime is prohibitively difficult. Besides, other physical obstacles appear, namely, the possible breaking of Lorentz symmetry. Even though it is true that this symmetry is broken by the finiteness of the defect (in the single-defect case), by shrinking the defect we approach Minkowski space. Therefore we can think of the parameter $b$ as controlling the breaking. 
The situation is more delicate with the gas of defects. Let us assume that this gas is distributed homogeneously over a lattice. The number of defects per volume will be constant in the rest frame. If we go to a boosted frame, the number of defects will increase in the direction of the boost, leading to a breaking of the Lorentz symmetry. A possible way out of this problem was studied in the context of causal sets  \cite{Dowker1,Bombelli,Henson}. The key point for distributing the defects is given by a Poisson process defined by the following probability
\be
P_n\left(V_4\right)=\frac{\left(\rho_f V_4\right)^n \exp \left(-\rho_f V_4 \right)}{n !}\label{Pois},
\ee  
where the constant parameter $\rho_f$ represents the density. This is interpreted as the probability of finding $n$ defects in the four-dimensional volume $V_4$. The average number of defects is provided by the mean of the Poisson distribution $\langle N(V_4)\rangle=\rho_f V_4$. Since the Lorentz transformations are volume-preserving diffeomorphisms, the number of defects remains invariant in all frames. However, in the situation we are interested,  the defects are distributed through spatial sections of the spacetime and therefore, the volume showing up in (\ref{Pois}) is $V_3$ rather than $V_4$. After a Lorentz transformation, the three-dimensional volume does not remain invariant, but it is modifed by a $\gamma\left(\equiv1/\sqrt{1-\beta^2}\right)$ factor, $d\tilde{V}_3=\gamma dV_3$. If we substitute $V_3$ in (\ref{Pois}), go to a boosted frame and expand in $\beta$ we get
\be
P_n(\tilde{V}_3)=P_n(V_3)\left(1+\frac{\rho_f V_3-n}{2}\beta^2+\mathcal{O}(\beta^4)\right)\label{Pois1},
\ee
where $\beta=v/c$ and the left hand side of (\ref{Pois1}) represents the probability in the boosted frame. The maximum of $P_n(V)$ is reached when $n=\langle N \rangle_{V_3}=\rho_f V_3$ (which is the quantity we are interested in what follows). At this value, the term of order $\beta^2$ is suppressed and even though the sprinkling is not Lorentz-invariant $P_n(V)$ receives the first correction at $\beta^4$ order. Therefore, although Lorentz symmetry is broken, the sprinkling of defects minimizes this breaking for the most probable distribution. In spite of this, we will see that the Lorentz symmetry can be accidentally restored in certain cases.


\subsection{$\mathcal{M}_4^{(\tau=0)}$ and $\mathcal{M}_4^{(\tau=1)}$ gas} 
\label{gass00}
 As we discussed before, the exact calculation of the Green functions in spacetimes with more than one defect is prohibitively difficult but under certain conditions one can obtain approximate solutions. Let us start with some considerations about the multiple defect manifolds.

\begin{figure}[h]
    \centering
    \includegraphics[width=0.4\textwidth]{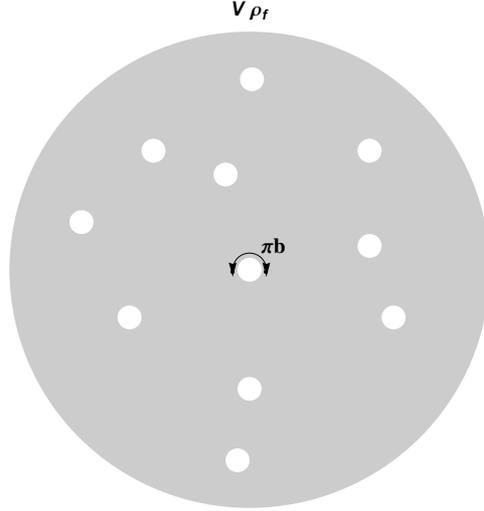}
    \caption{Gas of defects inside a spherical region of three-dimensional volume $V$.}
    \label{fig3}
\end{figure}

The spatial section of this new spacetime manifold can be written as the intersection of the single defect manifolds centered at $\{\vec{x}_1,...,\vec{x}_N\}$

\be
M^{(\tau)}_{\{\vec{x}_1,...,\vec{x}_N\};b}=\bigcap_{i=1}^N M_{\vec{x}_i;b}^{(\tau)}\label{mani1},
\ee
where the points $\{\vec{x}_1,...,\vec{x}_N\}$ are spread according to the Poisson distribution and the subscript $b$ stands for the size of the defects. Since in (\ref{mani1}) there is a finite number of defects the gas is confined to a compact region. For a infinite gas we can write formally
\be
\mathcal{M}_{4,\text{gas};b}^{(\tau)}=\mathbb{R}\times \lim_{N\rightarrow\infty}\left(M^{(\tau)}_{\{\vec{x}_1,...,\vec{x}_N\};b}\right).\label{manigas}
\ee

Let us take first the manifold $\mathcal{M}_4^{(\tau=0)}$. It is simple to see that, in the case of one defect at $\vec{x}_n$ the Green function (\ref{twot0}) takes the form
\be
G_\omega (\vec{x},\vec{x}')=G_{\omega,free} (\vec{x},\vec{x}')+\frac{b\vert \omega\vert^2}{4\pi}h_0^{(1)}(\vert \vec{x}-\vec{x}_n \vert\vert \omega\vert) h_0^{(1)}(\vert \vec{x}'-\vec{x}_n\vert\vert\omega\vert)+\mathcal{O}\left((\vert \omega\vert b)^2\right)\label{twot00}.
\ee

 As a consequence, the Green function corresponding to the multiple defect manifold can be written approximately as the sum of single defect manifolds
\be
G_\omega (\vec{x},\vec{x}')=G_{\omega,free} (\vec{x},\vec{x}')+\frac{b\vert \omega\vert^2}{4\pi}\sum_n h_0^{(1)}(\vert \vec{x}-\vec{x}_n \vert\vert \omega\vert) h_0^{(1)}(\vert \vec{x}'-\vec{x}_n\vert\vert\omega\vert)+\mathcal{O}\left((\vert \omega\vert b)^2\right)\label{twot11},
\ee
(here we are disregarding possible contributions from the correlations between couples of different defects). We will use the expression (\ref{twot11}) as our approximate Green function for the manifold (\ref{manigas}) with $\tau=0$. The regularity of the second term in (\ref{twot11}) in the coincidence limit suggests that it can be interpreted as a mass term. We can check it in the usual way. At linear order in $b$ the full Green function verifies
\be
\left(\nabla^2_x+\omega^2+m^2_{\tau=0}\right)G_\omega (\vec{x},\vec{x}')=\delta^{(3)}(\vec{x}-\vec{x}').
\ee

Using the properties of the Green function and taking into account that 
\be
\left(\nabla^2_x+\omega^2\right)\vert \omega\vert h_0^{(1)}(\vert \vec{x}-\vec{x}_n \vert\vert \omega\vert)=4\pi i\delta^{(3)}(\vec{x}-\vec{x}_n),
\ee
we obtain
\be
m_{\tau=0}^2G_{\omega,free} (\vec{x},\vec{x}')-ib\vert \omega \vert \sum_n \delta^{(3)}(\vec{x}-\vec{x}_n)h_0^{(1)}(\vert \vec{x}'-\vec{x}_n\vert\vert\omega\vert)+\mathcal{O}((b\vert\omega\vert)^2)=0\label{massgaszero}.
\ee

 Since we are distributing the defects according to a Poisson distribution, the number of defects grows with the three-dimensional volume. This allows us to replace the sum by the integral in (\ref{massgaszero}) according to the rule
 \be
 \sum_n\rightarrow \rho_f \int_{\mathbb{R}^3} d^3x\label{ass1},
 \ee
 which leads to 
 \be
 m_{\tau=0}^2=4\pi b \rho_f+...
 \ee
 
 This result can be confirmed easily in momentum space. After Fourier transforming the expression (\ref{twot11}) under the assumption (\ref{ass1}) we obtain
\bea
G(p)^{\tau=0}&=&\frac{1}{p^2+i\epsilon}-\frac{1}{p^2+i\epsilon} 4\pi b \rho_f \frac{1}{p^2+i\epsilon}=\nonumber\\
&=& \frac{1}{p^2-4\pi b\rho_f+i\epsilon}+...\label{gasfinprop}
\eea
 
 Therefore, the result of the interaction of the scalar field with the gas of defects is simply a generation of a mass term, $\frac{1}{2}m^2_{\tau=0}\phi^2$, proportional to the defect size and gas density. Note that  as a consequence of the sprinkling and at first order in $b$, the Lorentz symmetry is accidentally restored (the dispersion relation is the standard one for a massive scalar field). The explicit breaking at $b^2$ order can be seen from (\ref{twot0}) after Fourier transform. 

Under the same assumptions, the two-point function in coordinate space for $\mathcal{M}_{4,\text{gas};b}^{(\tau=1)}$ manifold is given by the following expression
\be
G_\omega (\vec{x},\vec{x}')=G_{\omega,free} (\vec{x},\vec{x}')+\frac{b^3\vert \omega\vert^4}{4\pi}\sum_n h_1^{(1)}(\vert \vec{x}-\vec{x}_n \vert\vert \omega\vert) h_1^{(1)}(\vert \vec{x}'-\vec{x}_n\vert\vert\omega\vert)\cos\gamma+...\label{twot12}
\ee
where $\gamma$ is the angle between $( \vec{x}-\vec{x}_n)$ and $( \vec{x}'-\vec{x}_n)$. We can now use the replacement (\ref{ass1}) and Fourier transform (\ref{twot12}) in $\Delta\vec{x}=(\vec{x}-\vec{x}')$. This corresponds to the expression for the propagator in momentum space. The full expression will be multiplied by the dimensionless quantity $b^3\rho_f$ and a simple dimensional analysis shows that no mass generation takes place. In momentum space we have
\bea
G(p)^{\tau=1}&=&\frac{1}{p^2+i\epsilon}-\frac{1}{p^2+i\epsilon}  b^3 \rho_f \vec{p}^2 \frac{1}{p^2+i\epsilon}=\nonumber\\
&=&\frac{1}{\omega^2-(1- b^3 \rho_f )\vec{p}^2+i\epsilon}+...
\eea

Therefore, the effect of the spacetime manifold $\mathcal{M}_4^{(\tau=1)}$ leads to a modification of the dispersion relation for scalar particles without mass generation.


\section{The photon propagator}
\label{propp}

In this section we discuss the photon propagation in a spacetime with one defect of the type $\tau=0$. We are interested in the following action
\be
S_{\text{photon}}=\int_{\mathcal{M}_4^{\tau=0}}d^4 x\left(\,-\frac{1}{4}F_{\mu\nu}F^{\mu\nu}-\frac{\lambda}{2}\left(\pa_\mu A^\mu\right)^2\right)\label{action},
\ee
where the last term in (\ref{action}) is the gauge fixing term. The free propagator (in Minkowski space) can be written as
\be
D^{\mu\nu}(x-y)=\langle 0\vert T A^\mu(x)A^\nu(y)\vert 0\rangle=\int\frac{d^4 k}{\left(2\pi\right)^4}e^{ik(x-y)}\left(\frac{\eta^{\mu\nu}}{k^2}-\frac{1-\lambda}{\lambda}\frac{k^\mu k^\nu}{(k^2)^2}\right).
\ee

In the Feynman gauge ($\lambda=1$) the propagator takes the simple form
\be
D^{\mu\nu}(x-y)=\eta^{\mu\nu}D(x-y)\label{prop},
\ee
where
\be
D(x-y)=\int\frac{d^4 k}{\left(2\pi\right)^4}\frac{e^{ik(x-y)}}{k^2}.
\ee

The free propagator (\ref{prop}) does not satisfy the boundary conditions for $\mathcal{M}_4^{\tau=0}$ manifold. We still have to impose boundary conditions on the defect. Let $\mathcal{S}$ be the surface of the defect. We have
\be
n^\mu F^\star_{\mu\nu}(x)\vert_{x\in\mathcal{S}}=0, \quad F^\star_{\mu\nu}=\frac{1}{2}\epsilon^{\mu\nu\rho\sigma}F_{\rho\sigma}\label{boundpho},
\ee
where $n^\mu$ in the outer normal vector to $\mathcal{S}$, $n^\mu=(0,\vec{n})$. These boundary conditions correspond to a vanishing normal component of the magnetic field and vanishing tangential components of the electric field 
\be
\vec{n}\wedge \vec{E}\vert_{x\in\mathcal{S}}=0,\quad \vec{n}\cdot \vec{B}\vert_{x\in\mathcal{S}}=0.
\ee

  It is important to note that the boundary conditions (\ref{boundpho}) are respected by the gauge transformations $\tilde{A}_\mu\rightarrow A_\mu+\pa_\mu\Lambda$ and hence if $A_\mu$ satisfies (\ref{boundpho}), $\tilde{A}_\mu$ also does. Following \cite{Bordag1,Bordag2} we rewrite the conditions (\ref{boundpho}) in the following form
\be
E_\mu^s A^\mu (s)\vert_{s\in\mathcal{S}}=0,\,\,s=1,2,\label{bound2}
\ee
where
\bea
E_\mu^1&=&\left(\begin{matrix}
    0   \\
     \vec{L}
\end{matrix}\right)\frac{1}{\sqrt{L^2}}, \label{pol1}\\
E_\mu^2&=&\left(\begin{matrix}
    L^2   \\
     i b \pa_0 \left(\vec{n}\times \vec{L}\right)
\end{matrix}\right)\frac{1}{\sqrt{L^2}\sqrt{-b^2\pa_0^2-L^2}},\label{pol2}
\eea
 $b$ is the defect radius and $\vec{L}=i \vec{x}\times \vec{\nabla}$. With this choice, the vectors $E_\mu^s$ fulfill the normalization condition $E_\mu^{s^\dagger}g^{\mu\nu}E^t_\nu=-\delta^{st}$. We can implement the conditions (\ref{bound2}) (or equivalently (\ref{boundpho})) by inserting a delta function in the generating functional of the Green functions
\be
Z\left(J\right)=N\int DA \prod_{\nu,x\in S}\delta\left(E_\mu^s A^\mu (s)\vert_{s\in\mathcal{S}}\right)\exp\lbrace i\lbrack S^{(0)}(A)+\int dx A_\mu J^\mu\rbrack\rbrace .
\ee

 Now by using a Fourier integral representation for the delta function and after diagonalization we obtain the photon propagator due to the presence of one defect (for details see \cite{Bordag1, Bordag2})
\be
D^c_{\mu\nu}(x,y)=D_{\mu\nu}(x-y)-\int_\mathcal{S} dz dz' D_\mu^{\,\,\,\rho}(x-z)E^s_\rho(z)K^{-1}_{st}(z,z')E^t_\sigma (z')D^\sigma_{\,\,\, \nu}(z'-y)\label{fullphot},
\ee
where
\be
 K^{st}(z,z')\equiv E^{s\dagger}_\mu(z)D^{\mu\nu} (z-z')E^t_\nu(z'),\quad  z,z'\in\mathcal{S}\label{heat},
\ee
 and $D_{\mu\nu}(x-y)$ is the free photon propagator. Since we want to calculate the correction term due to the spherical defect it is useful to expand the free propagators in spherical harmonics
\be
D(x-y)=\sum_{lm}\int_\mathbb{R} \frac{d \omega}{2\pi}e^{i \omega (x_0-y_0)}Y_{lm}(\theta_x,\phi_x)i\vert \omega\vert j_l(\vert\omega\vert r_\text{min})h_l^{(1)}(\vert\omega\vert r_{\text{max}})Y^\star_{lm}(\theta_y,\phi_y),  
\ee
where $j_l$ and $h^{(1)}_l$ are the spherical Bessel function of the first kind and the spherical Hankel function of the first kind respectively, $ r_\text{min}=\min (r_x,r_y)$ and $r_\text{max}=\max (r_x,r_y)$. We can determine the matrix $K^{st}$ from definition (\ref{heat}), invert the expression and expand for small defect radius. The first term in the expansion of the matrix $\left(K^{-1}\right)^{st}$ is particularly simple
\be
\left(K^{-1}\right)^{11}(z,z')=-\frac{9}{8\pi^2 b^3}\int d\omega \left( e^{i\omega (z_0-z_0')}\cos(z,z')+...\right) =\left(K^{-1}\right)^{22}(z,z') ,\label{k11}
\ee
and the non-diagonal terms vanish. Due to the structure of the $E_\mu^s(z)$ operators there are more leading terms in the expansion (\ref{k11}), but we do not display them here for the sake of simplicity. By inserting (\ref{k11}) in (\ref{fullphot}) taking into account (\ref{pol1}) and (\ref{pol2}) we get for the correction term in the space-space components
\be
D^b_{ij}(x,y)=\frac{b^3}{16\pi^2}\int d\omega e^{i\omega(x_0-y_0)} h_1^{(1)}(\vert \omega \vert \vert \vec{x} \vert)h_1^{(1)}(\vert \omega \vert  \vert \vec{y} \vert)\vert \omega\vert^4 \left(\delta_{i j} \hat{x}_k \hat{y}_k-\hat{x}_i \hat{y}_j\right)\label{cpho},
\ee
for the space-time components
\be
D^b_{0i}(x,y)=D^b_{i0}(x,y)=0,
\ee
and for the time-time component
\be
D^b_{00}(x,y)=\frac{b^3}{8\pi^2}\int d\omega e^{i\omega(x_0-y_0)} h_1^{(1)}(\vert \omega \vert \vert \vec{x} \vert)h_1^{(1)}(\vert \omega \vert  \vert \vec{y} \vert)\vert \omega\vert^4 \hat{x}_k \hat{y}_k,
\ee
where $\hat{x}_k, \hat{y}_k $ are the unit vectors in the direction of the spatial vectors $\vec{x},\vec{y}$ respectively. Our final expression for the photon propagator due to the interaction with one $\tau=0$ defect located at the origin (which corresponds to the free propagator $D_{\mu\nu} (x-y)$ plus the correction induced by the defects $D^b_{\mu\nu}(x,y)$) can be written as
\bea
D_{\mu\nu}^c(x,y)=D_{\mu\nu} (x-y)+\frac{b^3}{16\pi^2}\int d\omega e^{i\omega(x_0-y_0)} h_1^{(1)}(\vert \omega \vert \vert \vec{x} \vert)h_1^{(1)}(\vert \omega \vert  \vert \vec{y} \vert)\alpha_{\mu\nu}(\vec{x},\vec{y})\label{phofull},
\eea
where
\be
\alpha_{\mu\nu}=\left(\begin{matrix}
    2\hat{x}_k\hat{y}_k&0  \\
     0&\delta_{ij}\hat{x}_k\hat{y}_k-\hat{x}_i\hat{y}_j
\end{matrix}\right).
\ee

We end this section with a few words about the manifold $\mathbb{R}\times \mathcal{M}^{\tau=1}$. The boundary conditions for the field are now
\bea
n^\mu(x) F^{\star}_{\mu\nu}(x)\vert_{x\in\mathcal{S}}&=&n^\mu(-x) F^{\star}_{\mu\nu}(-x)\vert_{x\in\mathcal{S}},	\label{m11}\\
n^\mu(x) F_{\mu\nu}(x)\vert_{x\in\mathcal{S}}&=&-n^\mu(-x) F_{\mu\nu}(-x)\vert_{x\in\mathcal{S}}\label{m12}.
\eea

The relation (\ref{m11}) imposes conditions on the tangential components of the electric field ($\vec{E}$) and on the normal component of the magnetic ($\vec{B}$) field while (\ref{m12}) twists these conditions. Now we expand $\vec{E}$ and $\vec{B}$ in spherical harmonics ($Y_l^m$) and take the leading term (in the expansion for $b$ small). Since our defect is spherical we have $n^\mu(-x)=-n^\mu(x)$ and $Y^m_l(\pi-\theta,\pi+\varphi)=(-1)^l Y^m_l(\theta,\varphi)$. Therefore, at leading order the constraint (\ref{m11}) reduces to 
\be
n^\mu(x) F^{\star}_{\mu\nu}(x)\vert_{x\in\mathcal{S}}=0+\text{(higher orders)},
\ee
which corresponds to the boundary conditions for $\mathbb{R}\times \mathcal{M}^{\tau=0}$. The condition (\ref{m12}) is automatically satisfied. As a consequence, the propagators for both types of defects will coincide at leading order. From now on, the results obtained for $\mathbb{R}\times \mathcal{M}^{\tau=0}$ are understood as valid in both manifolds.


\subsection{$\mathcal{M}_4^{(\tau=0)}$ and $\mathcal{M}_4^{(\tau=1)}$ gas}
\label{sgas}

Our next step is the calculation of the photon propagator in a gas of defects. We cannot calculate the propagator by imposing boundary condition in each particular defect. Instead, and under some assumptions, we can determine an approximate propagator based of the single defect propagator. First, the spacetime manifold we are interested in can be defined formally as the intersection of single-defect manifold presented in Secs. 2 and 6.  If the defect is not located at the origin, but at the point $\vec{z}$, the correction part of  (\ref{phofull}) becomes
\be
D^b_{\mu\nu}(x,y;\vec{z})=D^b_{\mu\nu}(x_0,y_0,\vec{x}-\vec{z},\vec{y}-\vec{z}).
\ee

Note that the free part of the propagator is not modified.
Now, under the assumptions of Sec. 6 (we assume that the defects are distributed according to a Poisson process in such a way we can neglect the correlations between different defects) we can calculate approximately the photon propagator for the gas of defects \cite{ Klink2,Schreck,queiruga}). This implies that the characteristic distance defined by $\rho_f$ in (\ref{Pois}), $l_f=1/\rho_f^{1/3}$ must be much bigger that $b$. We can express the full propagator as a sum of individual propagators corresponding to the presence of single defects
\be
\sum_n D^b_{\mu\nu}(x_0,y_0,\vec{x}-\vec{z}_n,\vec{y}-\vec{z}_n)\label{corre}.
\ee

Instead of working directly with (\ref{corre}), we use the replacement (\ref{ass1}). From (\ref{Pois}) it is easy to see that the average number of defects in a region of volume $V$ is proportional to the volume, being the proportionality factor $\rho_f$.  We get therefore
\bea
D^b_{\mu\nu}(x,y;\rho_f)&=&\rho_f \int_{\mathbb{R}^3}d^3z D^b_{\mu\nu}(x_0,y_0,\vec{x}-\vec{z},\vec{y}-\vec{z}). \label{cpho}
\eea

Note that after using (\ref{ass1}) the integral is taken in $\mathbb{R}^3$ instead of the spatial section of $\mathcal{M}_{4,\text{gas}}^{(\tau)}$ . We have replaced the sum over the single defects by an integral. The density $\rho_f$ characterizes the continuous distribution of defects. Once we have the modified propagator (\ref{cpho}) our goal is to determine the modification of the dispersion relation as a result of the interaction with the spacetime foam. The modified photon propagator (\ref{cpho}) can be written in momentum space as follows (see Appendix A for useful integrals)
 \be
 D_{\mu\nu}^c(p)=\frac{\eta_{\mu\nu}}{p^2}-\frac{\rho_f b^3}{p^4}B_{\mu\nu}^b(p)\label{propmom},
 \ee
 where
 \bea
 B_{\mu\nu}&=&\vec{p}^2 k_{\mu\nu}-\omega_{\mu\rho}\omega_{\nu\sigma}p^\rho p^\sigma,\\
 k_{\mu\nu}&=&\text{diag}\left(1,1,1,1\right),\\
 \omega_{\mu\nu}&=&\text{diag}\left(0,1,1,1\right).
 \eea

After inverting (\ref{propmom})) we can write the field equations in momentum space as follows
 \be
 \left(\eta^{\mu\nu}p^2+\beta^{\mu\nu}(p) \right)A_\mu(p)=0\label{momeq},
\ee 
where
\be
 \beta_{\mu\nu}=\rho_f b^3\left(\vec{p}^2 k_{\mu\nu}-\omega_{\mu\rho}\omega_{\nu\sigma}p^\rho p^\sigma\right).
\ee

The diagonalization of the  polarization operator $\beta^{\mu\nu}$ leads to four eigenvectors and four eigenvalues. Two of these eigenvectors correspond to the scalar and longitudinal modes which are unphysical. The corresponding eigenvalues provide the dispersion relations
\bea
\omega^2&=&\vert \vec{p}\vert^2,\quad\quad \quad\text{longitudinal mode},\\
\omega^2&=&\vert \vec{p}\vert^2(1+2\rho_f b^3),\quad\text{scalar mode}.
\eea
 \noindent
 For the transverse modes we get
 \be
 \omega^2=\vert \vec{p}\vert^2(1-\rho_f b^3),\quad\text{transverse modes}.
 \ee

The dispersion relation for the longitudinal mode remains unchanged, while the scalar mode acquire a positive shift. This is not problematic since these modes are unphysical. However for the transverse modes, the physical ones, we get a negative shift. Higher-order corrections to this result are discussed in the following section.

\section{Summary and discussion}
\label{sum}

In this work we have described some aspects of particle propagation in spacetimes with certain types of defects. The presence of these defects modifies the Green functions. The correction terms to the Green functions have a regular behavior in the coincidence limit, when $x\rightarrow x'$. Moreover, if we consider single or finite defect manifolds these corrections vanish very far from the defect(s) leading therefore to the usual behavior of Lorentz-invariant theories. By means of the point splitting method we have calculated the modified vacuum energy-momentum tensor. We have found that the first correction to the tensor, in the $b$-expansion does not contribute, despite the fact that the Green function has a nonzero linear term. In particular, the vacuum energy is nonzero and negative close to the defect, and decays to zero like $1/r^6$ for $\mathcal{M}_4^{\tau=0}$ and like $1/r^7$ for $\mathcal{M}_4^{\tau=1}$. We have shown also that, due to the self-interaction term $\lambda\phi^4$, scalar particles can acquire a nonzero coordinate-dependent mass, which approaches zero far from the defect as $1/r^3$ for $\mathcal{M}_4^{\tau=0}$ and $1/r^5$ for $\mathcal{M}_4^{\tau=1}$.

In Secs. 4 and 5 we have considered single static defects embedded in flat space, and of course the foam would be formed by a certain density of defects distributed through the spacetime. The exact calculation of the Green functions in such a foam is prohibitively difficult, even for a small number of defects. In order to circumvent this problem we have approximated the Green functions for the gas of defects in terms of the Green function for the single defect manifold. We found that, due to the interaction with the defects, the scalar field acquires a mass which is proportional to the defect size and  gas density of the $\tau=0$ gas. Furthermore, we found that in this case the Lorentz symmetry is restored at linear order in $b$. However, the effect of the $\tau=1$ defects on the scalar field has different consequences. The first correction to the propagator appears at $b^3$ order and does not lead to a mass term but to a shift in the quadratic coefficient of the dispersion relation. Despite these differences, in both cases the usual Minkowski behavior is recovered in the limits $b\rightarrow 0$ (zero-size defects) and $\rho_f\rightarrow 0$ (zero-density gas). In the light of these results we can assure that the topological structure of a hypothetical spacetime foam leads to radically different physical consequences for scalar particles.
The situation with the photon is rather different. We have calculated the photon propagator due to the presence of static defects at leading order in the size parameter. We have analyzed two different topologies for the defects, and found that, at the order considered here, the photon propagators have the same form. This implies, in particular, that the photon cannot feel the difference between the topology of $\mathbb{D}^3\setminus\{p\}$ and $\mathbb{R}P^3\setminus\{p\}$ at this order. Of course, we expect different corrections beyond the leading order approximation. The transverse (and the only physical) modes exhibit the following dispersion relation
\be
\omega^2=\left(1- \rho_f b^3\right)\vert \vec{p}\vert^2.
\ee
This result is consistent with \cite{Mav1,Klink1,Klink2}. Moreover, our model does not exhibit birefringence as both polarization modes have the same dispersion relations, in agreement with experiments \cite{Gleiser, Fan}. Now, dimensional arguments and a simple analysis of the full propagator (\ref{fullphot}) allow us to obtain the subleading contribution to the modified dispersion relation. The expansion of the terms $D_\mu^{\,\rho}(x-z)D^{\sigma}_{\,\nu}(z'-y)$ and $E^s_\rho(z)E^t_\sigma (z')$ in (\ref{fullphot}) contains only even powers of $b$, while the expansion of $K_{st}^{-1}(z,z')$ only contains odd powers. As we have shown, the first term in the expansion is cubic in $b$, and the previous arguments show that the second must be of order $b^5$. Since $\rho_f$ can only appear linearly, a simple dimensional argument shows that the subleading modified dispersion relation has the form
\be
\omega^2=\left(1- \rho_f b^3\right)\vert \vec{p}\vert^2+\alpha^\tau \rho_f b^5\vert \vec{p}\vert^4,\label{dissub}
\ee  
where $\alpha^\tau$ is a dimensionless numerical coefficient which may depend on the defect topology (if zero, the first correction beyond the first order will have the form $\beta^\tau \rho_f b^7 \vert\vec{p} \vert^6$ but it does not change the discussion below). For completeness we can map the coefficients of the dispersion relation (\ref{dissub}) to coefficients of the Lorentz-violating Standard-Model Extension (SME) \cite{Koste1,Koste2}. The first correction corresponds to the $\tilde{\kappa}_{\text{tr}}$-type CPT-even Lorentz violating term
\be
\mathcal{L}^{\text{CPT-even}}=-\frac{1}{4}\left(k_F\right)_{\kappa\lambda\mu\nu}F^{\kappa\lambda}F^{\mu\nu},\label{sme1}
\ee
where
\bea
\left(k_F\right)_{\kappa\lambda\mu\nu}&=&\frac{1}{2}\left(\eta_{\kappa\mu}\tilde{\kappa}_{\nu\lambda}-\eta_{\kappa\nu}\tilde{\kappa}_{\lambda\mu}-\eta_{\lambda\mu}\tilde{\kappa}_{\kappa\nu}+\eta_{\lambda\nu}\tilde{\kappa}_{\kappa\mu}\right),\\
\tilde{\kappa}_{\mu\nu}&=&\frac{3}{2}\tilde{\kappa}_{\text{tr}}\text{diag}\left(1,\frac{1}{3},\frac{1}{3},\frac{1}{3}\right).
\eea

This gives the correspondence
\be
\rho_f b^3=\frac{2\tilde{\kappa}_{\text{tr}}}{1+\tilde{\kappa}_{\text{tr}}}.\label{kappa1}
\ee

The subleading correction (if $\alpha^\tau\neq 0$) corresponds to a part of the nonminimal photon-sector of the SME. For example we can take
\be
\mathcal{L}^{\text{CPT-even}}_{\text{nonminimal}}=-\frac{1}{4}F^{\kappa\lambda}\left(\hat{k}_F\right)_{\kappa\lambda\mu\nu}F^{\mu\nu}\label{sme2},
\ee
where
\bea
\left(\hat{k}_F\right)_{\kappa\lambda\mu\nu}&=&\frac{1}{2}\left(\eta_{\kappa\mu}\tilde{\kappa}'_{\nu\lambda}-\eta_{\kappa\nu}\tilde{\kappa}'_{\lambda\mu}-\eta_{\lambda\mu}\tilde{\kappa}'_{\kappa\nu}+\eta_{\lambda\nu}\tilde{\kappa}'_{\kappa\mu}\right)\tilde{\eta}^{\alpha_1\alpha_2}\pa_{\alpha_1}\pa_{\alpha_2},\\
\tilde{\kappa}'_{\mu\nu}&=&\frac{3}{2}\tilde{\kappa}_{\text{tr}}'\text{diag}\left(1,\frac{1}{3},\frac{1}{3},\frac{1}{3}\right),\\
\tilde{\eta}^{\alpha_1\alpha_2}&=&\text{diag}\left(0,1,1,1\right).
\eea

If we consider together (\ref{sme1}) and (\ref{sme2}) we obtain a new correspondence
\be
\alpha^\tau \rho_f b^5=\frac{2\tilde{\kappa}_{\text{tr}}'}{\left(1+\tilde{\kappa}_{\text{tr}}\right)^2}.
\ee

  We first see that the dispersion relation does not contain cubic terms in $\vert \vec{p}\vert$, consistent with the rotational invariance \cite{Lehnert} and besides, is in agreement with the results of \cite{Klink2} for the photon propagation. Now,  from the density $\rho_f$ we can define a characteristic distance between defects $l_{f}=1/\rho_f^{1/3}$ as discussed in Sec. 7. In analogy with \cite{Klink2}, we use the bounds obtained in \cite{Koste3} for the coefficient $\tilde{\kappa}_{\text{tr}}$ for a dispersion relations of type (\ref{dissub}) and with the correspondence (\ref{kappa1}). If $b$ is assumed to be of the order of the Planck length $b\sim l_{P}$ then $l_f\gtrsim 1.5\times 10^6\, l_P$ which is consistent with our approximation disregarding correlations between different defects (\ref{corre}) and rules out the possibility of a single-scale foam with $l_f\sim b$. 
The topologies considered here may only affect the numerical coefficient $\alpha^\tau$ and therefore they lead to the same conclusions. But, of course, other choices for the topological structure like mixtures of topologies or time-dependent defects, might change the dispersion relations leading to other bounds for $l_f$ and different correspondences to the SME.
 These generalizations are under current investigation.

{\bf Acknowledgements.}- It is a pleasure to thank F.R. Klinkhamer for extensive discussions. 
The author also thanks M. Mole for his careful reading of the manuscript.


\appendix

\section{Useful integrals}
\label{a1}

We begin with the following integral
\be
\int d^3 x\frac{e^{-i\vec{p}\cdot\vec{x}}}{4\pi \vert \vec{x}\vert }e^{i\vert \omega\vert\vert \vec{x}\vert}=\frac{1}{\vert \vec{p}\vert^2-\omega^2}\label{int1}.
\ee

If we differentiate with respect to $p_i$ on both sides we obtain 
\be
\int d^3 x\frac{e^{-i\vec{p}\cdot \vec{x}}}{4\pi \vert \vec{x}\vert }e^{i\vert \omega\vert \vert \vec{x}\vert}x_i=\frac{-2i p_i}{\left(\vert \vec{p}\vert^2-\omega^2\right)^2},
\ee
where $x_i=\vert \vec{x}\vert \hat{x}_i$ and $\hat{x}_i$ the unit vector in the i-th direction.  In most of our calculations the function in the integrand is the spherical Hankel function, $h_1^{(1)}(z)$
\be
h_1^{(1)}(z)=-e^{iz}\frac{z+i}{z^2}.
\ee 
 Thus, it will be useful to have a closed expression for the integral (\ref{int1}) for arbitrary powers of $ \vert \vec{x}\vert$ in the denominator. Formally this integral can be written as follows (see for example \cite{Gradshtein}),
\be
\int d^3 x\frac{e^{-i \vec{p}\cdot \vec{x}}}{4\pi \vert \vec{x}\vert^n }e^{i\vert \omega\vert \vert\bar{x}\vert}=\frac{i \Gamma (3-n) \left(\frac{(i (\vert \vec{p}\vert-\vert\omega\vert ))^n}{(\vert    \vec{p}\vert-\vert\omega\vert
   )^2}-\frac{(-i (\vert \vec{p}\vert+\vert\omega\vert ))^n}{(\vert \vec{p}\vert+\vert\omega\vert )^2}\right)}{2 (n-2)
 \vert  \vec{p}\vert}.
\ee

For our calculation we are interested in the $n=2,3$ integrals, and despite the fact that this integral in divergent for $n\geq 3$ (due to the poles of the $\Gamma$ function), its derivatives with respect to $p_i$ are well-behaved. We have
\be
I=\int d^3 x\frac{e^{-i \vec{p}\cdot\vec{x}}}{4\pi \vert \vec{x}\vert }h_1^{(1)}(\vert \omega\vert \vert \vec{x}\vert)\vert \omega\vert^2 \hat{x}_i=I_1+I_2
\ee
where
\bea
I_1&=&-\vert\omega\vert\int d^3 x\frac{e^{-i \vec{p} \cdot\vec{x}}}{4\pi \vert \vec{x}\vert^2 }e^{i\vert \omega\vert \vert\bar{x}\vert}	 \hat{x}_i\\
I_2&=&-i\int d^3 x\frac{e^{-i \vec{p}\cdot \vec{x}}}{4\pi \vert \vec{x}\vert^3 }e^{i\vert \omega\vert \vert \vec{x}\vert}	 \hat{x}_i.
\eea
After integration we get
\bea
I_1&=&-\frac{i p_i \vert \omega\vert}{2\vert \vec{p}\vert^3 (\vert \vec{p}\vert^2-\omega^2)}\left(-2\vert \vec{p}\vert \vert \omega\vert +(\vert \vec{p}\vert^2-\omega^2)\log\frac{\vert\omega\vert-\vert \vec{p}\vert}{\vert\omega\vert+\vert \vec{p}\vert}\right)\\
I_2&=&\frac{i p_i }{2\vert \vec{p}\vert^3 }\left(2\vert \vec{p}\vert +\vert\omega\vert\log\frac{\vert\omega\vert-\vert \vec{p}\vert}{\vert\omega\vert+\vert \vec{p}\vert}\right).
\eea
The logarithmic contributions disappear in the combination $I_1+I_2$ and finally we obtain
\be
I=-i\frac{p_i}{\omega^2-\vert \vec{p}\vert^2}.
\ee






\section{Jump condition}
We can understand the jump condition (\ref{b3}) in the following way. Let us take the radial part of the Laplacian after Fourier transform
\be
\left(\omega^2+\frac{\pa^2}{\pa r^2}+\frac{2}{r}\frac{\pa}{\pa r}\right)G_\omega (r,r',\gamma)=\frac{1}{r^2\sin(\theta)}\delta(r-r')\delta(\theta-\theta')\delta(\phi-\phi')\label{eq1},
\ee 
where $G_\omega (r,r',\gamma)$ is a Green function in the notation used above. If we take the integral $\lim_{\epsilon\rightarrow 0}\int_{r'-\epsilon}^{r'+\epsilon}$ in the l.h.s. the zeroth order operator ($\omega^2$) and the first order operator ($\pa/\pa r$) vanish in the limit $\epsilon\rightarrow 0$ due to the continuity of $G_\omega (r,r',\gamma)$ at $r=r'$. For the second order operator we get
\be
\lim_{\epsilon\rightarrow 0}\int_{r'-\epsilon}^{r'+\epsilon} r^2\frac{\pa^2}{\pa r^2}G_\omega (r,r',\gamma)=\left(r^2\pa_r G_\omega (r,r',\gamma)\vert_{r=r'^+}-r^2\pa_r G_\omega (r,r',\gamma)\vert_{r=r'^-} \right),
\ee
where the first term in the r.h.s in evaluated at $r>r'$ and the second at $r<r'$. Now substitute the expressions for $G_\omega (r,r',\gamma)$, multiply this expression by $P_m(\cos\gamma)$ and integrate in both sides of (\ref{eq1}) using
\be
\int d\theta d\phi\sin\theta P_m(\cos\gamma) P_n(\cos\gamma)=\delta_{m n}\frac{4\pi}{2 n +1}.
\ee
This gives
\be
\frac{r'^2}{\sqrt{r'}}\left(a_n H_{n+1/2}^{(1)'}-b_n H_{n+1/2}^{(1)'}-c_n H_{n+1/2}^{(2)'}\right)\frac{4\pi}{2 n +1}=1\label{jump}.
\ee
Therefore, the discontinuity of the derivative (jump condition) appears as a consequence of the Dirac delta in the definition of the Green function with the factors obtained in the equation above.

\end{document}